spepping $
LaTeX2.09
\documentclass[fleqn,12pt,twoside]{article}
\usepackage{espcrc1}
\usepackage{graphicx}
\usepackage[figuresright]{rotating}

\newcommand{\AmS}{{\protect\the\textfont2
  A\kern-.1667em\lower.5ex\hbox{M}\kern-.125emS}}
\input{tcilatex}

\begin{document}

\title{Infrared Sensitivity in Damping Rate for Very 
Soft Moving Fermions in
Finite Temperature QED}
\author{A. Abada%
\address{D\'{e}partement de Physique, Ecole Normale 
Sup\'{e}rieure  \\
BP 92 Vieux Kouba, 16050 Alger, Algeria}\thanks{\texttt
{abada@wissal.dz}}
and K. Bouakaz\addressmark\thanks{\texttt
{bouakazk@caramail.com}}}
\maketitle

\begin{abstract}
We calculate the fermion damping rate to second order in 
powers of the
external momentum $p$ in the context of QED at finite 
temperature using the
hard-thermal-loop (HTL) summation scheme. We find that 
the coefficient of
order $p^{2}$ is divergent in the infrared whereas the 
two others are
finite. This result suggests that the htl-based 
pertubation is infrared
sensitive at next-to-leading order.
\end{abstract}

\bigskip

At high temperature, perturbative calculations based on 
bare propagators and
vertices lead to gauge dependent and infrared divergent 
results \cite{sil}.
This problem has been solved within the hard thermal 
loop (HTL) summation
technique \cite{Pis,Bra B337}. The basic idea is to 
distinguish between
momentum scales $T$ (hard) and $eT$ (soft), where $e$ is 
the QED coupling
constant. Using this method, the explicit calculation of 
the zero-momentum
transverse gluon damping rate $\gamma _{t}\left( 0
\right) $ to
next-to-leading order was performed in \cite{gamt} and 
was shown to be
finite, positive and independent of the gauge. A similar 
computation for the
damping rate of a quark at rest was carried 
independently in \cite{kobes}
and\ \cite{braaten (quarks)}.

However, recent work shows that the damping rates for 
very soft moving
excitations in hot QCD are infrared sensitive. Indeed, 
an explicit
calculation of the longitudinal gluon damping rate 
$\gamma _{l}\left(
0\right) $ shows that this latter is different from 
$\gamma _{t}\left(
0\right) $ and infrared divergent \cite{AAB,AA}. Also, 
the calculation of
the transverse gluon damping rate with very soft 
momentum $p$ is discussed
in \cite{AAT} and presented in the form $\gamma _{t}\left
( p\right)
=a_{t0}+a_{t1}\left( \frac{p}{m_{g}}\right) ^{2}+...$, 
where $m_{g}$ is the
soft gluon mass. The zeroth-order coefficient $a_{t0}$ 
is finite and equal
to $\gamma _{t}\left( 0\right) $ found in \cite{gamt}, 
but the second-order
coefficient is infrared divergent. Finally, the 
calculation of the damping
rate for quarks to second order in the external momentum 
discussed in \cite%
{ABD} suggests also that the second-order coefficient is 
divergent whereas
the zeroth-order and the first-order ones are finite.

In this work, we calculate the damping rates of very 
soft fermions at
next-to-leading order in finite temperature QED. We use 
the imaginary-time
formalism with Euclidean momenta $P^{\mu }=(p_{0},
\overrightarrow{p})$ so
that $P^{2}=p_{0}^{2}+p^{2}$ and $p=\left\vert 
\overrightarrow{p}\right\vert 
$. We follow the conventions of \cite{Lebellac}. The 
fermion propagator is
obtained by summing the hard-thermal-loop fermion self-
energy diagrams. It
is defined by $^{\ast }\Delta _{f}^{-1}\left( P\right) 
=P\hspace{-8pt}%
/-\delta \Sigma \left( P\right) $, where $\delta \Sigma 
$ is the HTL in the
fermion self-energy. The effective fermion propagator 
can be decomposed into
the two helicity eigenstates: 
\begin{equation}
^{\ast }\Delta _{F}\left( P\right) =-\frac{1}{2}\left[ 
\left( \gamma ^{0}-i%
\mathbf{\gamma }.\widehat{\mathbf{p}}\right) \Delta _{+}
\left( P\right)
+\left( \gamma ^{0}+i\mathbf{\gamma }.\widehat{\mathbf
{p}}\right) \Delta
_{-}\left( P\right) \right] ,  \label{effe-prop-f}
\end{equation}%
where $\Delta _{\pm }(P)=\left( ip_{0}\mp p-\frac{m_{f}^
{2}}{2p}\left[
\left( 1\mp \frac{ip_{0}}{p}\right) \ln \dfrac{ip_{0}+p}
{ip_{0}-p}\right]
\pm 2\right) ^{-1}$; $m_{f}=eT/\sqrt{8}$ is the fermion 
thermal mass. The
poles of $\Delta _{\pm }(-i\omega ,\mathbf{p})$ define 
the dispersion
relation $\omega _{\pm }(p)$ for fermions to leading 
order in $e$. In
Covariant gauge, the effective photon propagator is: 
\begin{equation}
^{\ast }\Delta _{\mu \nu }\left( K\right) =\frac{1}{G+K^
{2}}P_{\mu \nu }^{T}+%
\frac{1}{F+K^{2}}P_{\mu \nu }^{L}+\frac{\xi }{K^{2}}\frac
{K_{\mu }K_{\nu }}{%
K^{2}},  \label{effe-prop-p}
\end{equation}%
where $P_{\mu \nu }^{T},$ $P_{\mu \nu }^{L}$ are the 
transverse and
longitudinal projectors respectively and $F=\frac{2m^{2}
K^{2}}{k^{2}}\left(
1-\frac{ik_{0}}{k}Q_{0}\left( \frac{ik_{0}}{k}\right) 
\right) ,G=\frac{%
m^{2}ik_{0}}{k}\left( \left( 1-\left( \frac{ik_{0}}{k}
\right) ^{2}\right)
Q_{0}\left( \frac{ik_{0}}{k}\right) +\frac{ik_{0}}{k}
\right) ,$ with $m=eT/%
\sqrt{6}$ , the photon thermal mass.

The effective vertices $^{\ast }\Gamma $ are the sum of 
the tree amplitudes $%
\Gamma $ and the hard thermal loops $\delta \Gamma $. 
The effective
fermion-photon vertex and the effective two fermion-two 
photon vertex are
given by:%
\begin{equation}
^{\ast }\Gamma ^{\mu }(P_{1},P_{2})=\gamma ^{\mu }-m_{f}^
{2}\int \frac{%
d\Omega _{s}}{4\pi }\frac{S^{\mu }S\hspace{-0.6614pc}/}
{P_{1}S\,P_{2}S},
\label{effe-ver1}
\end{equation}%
\begin{equation}
^{\ast }\widetilde{\Gamma }^{\mu \nu }(P_{1},P_{2};Q)=-m_
{f}^{2}\int \frac{%
d\Omega _{s}}{4\pi }\left[ \frac{1}{P_{1}S}+\frac{1}{P_
{2}S}\right] \frac{%
S^{\mu }S^{\nu }S\hspace{-0.6589pc}/}{\left( P_{1}
+Q\right) \hspace{-2pt}%
S\left( P_{2}-Q\right) \hspace{-2pt}S}\,.  \label{effe-
ver2}
\end{equation}%
The damping rates for fermions to second order in their 
very soft momentum $%
p $ write: 
\begin{equation}
\gamma _{\pm }\left( p\right) =\left. \frac{1}{2}\left[ 1
\pm \frac{2}{3}%
\frac{p}{m_{f}}-\frac{2}{9}\left( \frac{p}{m_{f}}\right) 
^{2}+\dots \right]
\,\mathrm{Im\,}^{\ast }f_{\pm }\left( -i\omega ,p\right) 
\right\vert
_{\omega =\omega _{\pm }+i0^{+}}\,,
\end{equation}%
where $^{\ast }f_{\pm }=\,^{\ast }D_{0}\mp \,^{\ast }D_
{s}$ and $\,^{\ast
}D_{0},\,^{\ast }D_{s}$ are the components of the 
fermion effective
self-energy. Therefore, the calculation of $\gamma _{\pm 
}\left( p\right) $
amounts to that of the imaginary part of the fermion 
effective self-energy.

Up to next-to-leading-order, the fermion self-energy is 
given in imaginary
time formalism by: 
\begin{equation}
\hspace{-0.7cm}^{\ast }\Sigma \left( P\right) =\hspace{-
3pt}\hspace{-2pt}%
-e^{2}\mathrm{Tr}_{\mathrm{soft}}\hspace{-2pt}\left[ ^
{\ast }\Gamma ^{\mu
}\left( P\hspace{-2pt},Q\right) \,\hspace{-2pt}\hspace{-
2pt}^{\ast }\Delta
_{F}\hspace{-2pt}\left( Q\right) \,\hspace{-2pt}^{\ast }
\Gamma ^{\nu }%
\hspace{-2pt}\left( P,Q\,\right) ^{\ast }\Delta _{\mu 
\nu }\left( \hspace{%
-1pt}K\hspace{-1pt}\right) \hspace{-3pt}+^{\ast }\Gamma ^
{\mu \nu }\hspace{%
-2pt}\left( P,\hspace{-2pt}-P;Q\,,\hspace{-2pt}-Q\right) 
^{\ast }\Delta
_{\mu \nu }\hspace{-2pt}\left( K\right) \right] ,
\end{equation}

where $K$ is the internal soft loop momentum, $Q=P-K$ 
and $\mathrm{Tr}\equiv
T\dsum \int \dfrac{d^{3}k}{(2\pi )^{3}}$. In this latter 
sum, only soft
values of $k$ are allowed in the integrals. Using the 
expressions of the
effective propagators (\ref{effe-prop-f}, \ref{effe-prop-
p}) and the
effective vertices (\ref{effe-ver1}, \ref{effe-ver2}), 
we see that there are
nine contributions to $^{\ast }\Sigma $. Details on how 
to handle these is
given in \cite{ABD}. We find that the damping rates can 
be put in the
following form: 
\begin{equation}
\gamma _{\pm }\left( p\right) =-\frac{e^{2}T}{8\pi }\left
[ a_{0}\pm \frac{p}{%
3}a_{1}+\frac{p^{2}}{9}a_{2}+\dots \right] ,
\end{equation}%
with: 
\begin{eqnarray}
a_{i} &=&\int_{\eta }^{\infty }dk\int_{-\infty }^{+
\infty }d\omega
\int_{-\infty }^{+\infty }\frac{d\omega ^{\prime }}
{\omega ^{\prime }}\left[
f_{i}\left( \omega ,\omega ^{\prime };k\right) -\,f_{i-
1,i\geq 1}\left(
\omega ,\omega ^{\prime };k\right) \,\partial _{\omega }
\right.  \nonumber \\
&&\left. +f_{i-2,i\geq 2}\left( \omega ,\omega ^{\prime 
};k\right) \left[
-3\,\partial _{\omega }+\partial _{\omega }^{2}\right] 
\right] \delta \,; \\
f_{0}\left( \omega ,\omega ^{\prime };k\right) &=&\sum_
{\varepsilon =\pm } 
\left[ -k^{2}\left( 1-\varepsilon k+\omega \right) ^{2}
\rho _{\varepsilon
}\rho _{l}^{\prime }+\frac{1}{2}\left( 1+2\varepsilon 
k+k^{2}-\omega
^{2}\right) ^{2}\rho _{\varepsilon }\rho _{t}^{\prime }
\right] \hspace{-3pt}
\nonumber \\
&&+\frac{1}{k}\left( k^{2}-\omega ^{2}\right) \rho _{0}
\rho _{t}^{\prime },
\\
f_{1}\left( \omega ,\omega ^{\prime };k\right) &=&\sum_
{\varepsilon =\pm } 
\left[ 2k^{2}\left( -1+k^{2}-2\varepsilon k\omega +
\omega ^{2}\right) \rho
_{\varepsilon }\rho _{l}^{\prime }+\left( -\frac{2
\varepsilon }{k}%
-3+2\varepsilon k+4k^{2}-k^{4}\right. \right.  \nonumber 
\\
&&\ \hspace{-1.1in}\left. -\left( 2+4\varepsilon k+2k^{2}
\right) \omega
+\left( \frac{4\varepsilon }{k}+4+2\varepsilon k+2k^{2}
\right) \omega
^{2}+2\omega ^{3}-\left( \frac{2\varepsilon }{k}+1
\right) \omega ^{4}\right]
\rho _{\varepsilon }\rho _{t}^{\prime }  \nonumber \\
&&\ \ \hspace{-1.1in}+\left. \hspace{-3pt}\varepsilon k^
{2}\hspace{-2pt}%
\left( 1\hspace{-2pt}-\varepsilon k+\hspace{-2pt}\omega 
\right) ^{2}\hspace{%
-3pt}\rho _{\varepsilon }\partial _{k}\rho _{l}^{\prime }
+\hspace{-1pt}%
\left( \frac{\varepsilon }{2}+2k+3\varepsilon k^{2}+2k^
{3}+\frac{\varepsilon 
}{2}k^{4}-\left( \varepsilon +2k+\varepsilon k^{2}
\right) \allowbreak \omega
^{2}+\frac{\varepsilon }{2}\omega ^{4}\allowbreak 
\right) \hspace{-3pt}\rho
_{\varepsilon }\partial _{k}\rho _{t}^{\prime }\right]  
\nonumber \\
&&\ \ \hspace{-1.1in}-\frac{2}{k}\left( k^{2}-\omega ^{2}
+2\frac{\omega ^{3}%
}{k^{2}}\right) \hspace{-3pt}\rho _{0}\rho _{t}^{\prime 
}-\hspace{-3pt}%
2k^{2}\epsilon \left( \omega \right) \delta \left( 
\omega ^{2}-k^{2}\right) 
\hspace{-3pt}\rho _{l}^{\prime }+\hspace{-3pt}\dfrac
{\omega }{k^{2}}\hspace{%
-3pt}\left( \omega ^{2}-k^{2}\right) \hspace{-3pt}\rho _
{0}\partial _{k}\rho
_{t}^{\prime }+2\omega \rho _{0}\partial _{k}\rho _{l}^
{\prime }\,;
\end{eqnarray}%
\begin{eqnarray}
f_{2}\left( \omega ,\omega ^{\prime };k\right) &=&\sum_
{\varepsilon =\pm }%
\hspace{-3pt}\left[ \left( -\frac{9}{2}-k^{2}-6
\varepsilon k^{3}-\frac{1}{2}%
k^{4}-\left( 6\varepsilon k-6k^{2}+2\varepsilon k^{3}
\right) \allowbreak
\omega +\left( 9+k^{2}\right) \omega ^{2}+6\varepsilon 
k\omega ^{3}\right.
\allowbreak \right.  \nonumber \\
&&\hspace{-1.03in}-\left. \frac{9}{2}\omega ^{4}\hspace{-
3pt}\right) \hspace{%
-3pt}\rho _{\varepsilon }\rho _{l}^{\prime }+\hspace{-
2pt}\hspace{-2pt}%
\hspace{-3pt}\left( \frac{9}{2k^{2}}-\hspace{-2pt}\frac
{14\varepsilon }{k}-%
\hspace{-2pt}\frac{8}{3}+\hspace{-3pt}4\varepsilon k-
\hspace{-2pt}\frac{19}{2%
}k^{2}\hspace{-2pt}-\hspace{-3pt}6\varepsilon k^{3}
\hspace{-2pt}+\hspace{-2pt%
}k^{4}\allowbreak \hspace{-2pt}+\hspace{-3pt}\left( \frac
{-3}{k^{2}}+\frac{%
25\varepsilon }{k}\hspace{-2pt}-\hspace{-2pt}10\hspace{-
2pt}+\hspace{-3pt}%
6\varepsilon k\hspace{-2pt}+\hspace{-2pt}9k^{2}\hspace{-
2pt}-\hspace{-3pt}%
3\varepsilon k^{3}\hspace{-2pt}\right) \omega \right.  
\nonumber \\
&&\hspace{-1.03in}+\left( -\frac{6}{k^{2}}+\frac{2
\varepsilon }{k}%
+23-6\varepsilon k+k^{2}\right) \allowbreak \omega ^{2}+
\left( \frac{6}{k^{2}%
}-\frac{22\varepsilon }{k}-6+6\varepsilon k\right) 
\allowbreak \omega
^{3}+\left( -\frac{3}{2k^{2}}+\frac{12\varepsilon }{k}-5
\right) \allowbreak
\omega ^{4}  \nonumber \\
&&\hspace{-1.03in}-\left. \left( \frac{3}{k^{2}}+\frac{3
\varepsilon }{k}%
\right) \allowbreak \omega ^{5}+\frac{3}{k^{2}}
\allowbreak \omega
^{6}\right) \rho _{\varepsilon }\rho _{t}^{\prime }-
k\left( 9-14\varepsilon
k+5k^{2}+\left( 12-2\varepsilon k-6k^{2}\right) \omega -
\left(
3-12\varepsilon k\right) \omega ^{2}\right.  \nonumber \\
&&\hspace{-1.03in}-\left. 6\omega ^{3}\right) \rho _
{\varepsilon }\partial
_{k}\rho _{l}^{\prime }+\left( \frac{3}{2k}+4\varepsilon 
+7k+8\varepsilon
k^{2}+\frac{7}{2}k^{3}-\left( \frac{3}{k}+6\varepsilon 
+6k+6\varepsilon
k^{2}+3k^{3}\right) \omega \right.  \nonumber \\
&&\hspace{-1.03in}-\left. \left( \hspace{-2pt}\frac{3}{k}
+4\varepsilon
+5k\right) \omega ^{2}+\hspace{-2pt}\hspace{-2pt}\left( 
\frac{6}{k}+\hspace{%
-2pt}6\varepsilon +6k\right) \omega ^{3}+\frac{3}{2k}
\omega ^{4}\hspace{-2pt}%
-\frac{3}{k}\omega ^{5}\right) \rho _{\varepsilon }
\partial _{k}\rho
_{t}^{\prime }-\frac{3}{2}k^{2}\left( 1-\varepsilon k+
\omega \right)
^{2}\rho _{\varepsilon }\partial _{k}^{2}\rho _{l}^
{\prime }  \nonumber \\
&&\hspace{-1.03in}+\left. \left( \frac{3}{4}+3
\varepsilon k+\frac{9}{2}%
k^{2}+3\varepsilon k^{3}+\frac{3}{4}k^{4}-\frac{3}{2}
\left( 1+2\varepsilon
k+k^{2}\right) \omega ^{2}+\frac{3}{4}\omega ^{4}\right) 
\rho _{\varepsilon
}\partial _{k}^{2}\rho _{t}^{\prime }\right] \hspace{-
3pt}-\frac{3}{k}\left(
k^{2}-\omega ^{2}\right) \rho _{0}\rho _{l}^{\prime }  
\nonumber \\
&&\hspace{-1.03in}+\left( \frac{3}{k}+2k+\frac{6}{k}
\omega -\left( \frac{15}{%
k^{3}}+\frac{2}{k}\right) \omega ^{2}+\frac{18}{k^{3}}
\omega ^{3}\right)
\rho _{0}\rho _{t}^{\prime }+\left( 6-12\omega \right) 
\rho _{0}\partial
_{k}\rho _{l}^{\prime }  \nonumber \\
&&\hspace{-1.03in}+\left( -3+6k\omega +\frac{3}{k^{2}}
\omega ^{2}-\frac{6}{k}%
\omega ^{3}\right) \rho _{0}\partial _{k}\rho _{t}^
{\prime }+3k\rho
_{0}\partial _{k}^{2}\rho _{l}^{\prime }-\frac{3}{2k}
\left( k^{2}-\omega
^{2}\right) \rho _{0}\partial _{k}^{2}\rho _{t}^{\prime 
}  \nonumber \\
&&\hspace{-1.03in}+12k^{2}\epsilon \left( \omega \right) 
\delta \left(
\omega ^{2}-k^{2}\right) \rho _{l}^{\prime }-6\left\vert 
\omega \right\vert
\delta \left( \omega ^{2}-k^{2}\right) \rho _{t}^{\prime 
}-6k^{2}\left\vert
\omega \right\vert \,\partial _{\omega ^{2}}\delta \left
( \omega
^{2}-k^{2}\right) \rho _{l}^{\prime }.
\end{eqnarray}%
In the above expression, $\eta $ is an infrared cutoff 
of the order of the
magnetic scale. $\delta =\delta \left( 1-\omega -\omega ^
{\prime }\right) $
and $\partial _{\omega }\equiv \frac{\partial }{\partial 
\omega }$; $\rho
_{\epsilon }=\rho _{\epsilon }\left( k,\omega \right) $ 
and $\rho
_{t,l}^{\prime }=\rho _{t,l}\left( k,\omega ^{\prime }
\right) $ are the
spectral densities of the effective propagators $^{\ast }
\Delta _{\epsilon }$
and $^{\ast }\Delta _{t,l}$ respectively \cite
{Lebellac,rho}. Also, $\rho
_{0}=-\frac{1}{2}\Theta \left( k^{2}-\omega ^{2}\right) 
$ and $\epsilon
\left( \omega \right) $ is the sign function. Note that 
for convenience, we
have set $m_{f}=1$ in the above formula.

What remains to do is the numerical evaluation of $a_{0}
$, $a_{1}$ and $a_{2}
$. For this purpose, we have to perform the integrals 
over the frequencies $%
\omega $ and $\omega ^{\prime }$, and then over the 
momentum $k$. This
necessitates the explicit use of the expressions of the 
spectral functions
of the different quantities involved \cite{Lebellac} and 
a delicate
extraction of potential infrared divergences. All 
relevant details will be
given elsewhere. We report here on the final result, 
which writes as:

\begin{equation}
\gamma _{\pm }\left( p\right) =\frac{e^{2}T}{16\pi }\left
[ 5.4253\mp
0.556\,6p+\left( -13.\,\allowbreak 3136\ln \eta +10.\,
\allowbreak
432\,7\right) p^{2}+...\right] .  \label{val damping}
\end{equation}

We have calculated the damping rates of fermions in 
finite-temperature QED
to next-to-leading order in HTL-summed perturbation. We 
have found that the
coefficients of zeroth order and order $p$ are infrared 
safe whereas the
third one is logarithmically divergent. This result 
indicates that the
HTL-summation scheme is perhaps not complete at this 
order.

\end{document}